\documentclass[fleqn,twoside]{article}
\usepackage{espcrc2}
\usepackage{epsf}
\input{psfig}

\usepackage{graphicx}
\newcommand{\AmS}{{\protect\the\textfont2
     A\kern-.1667em\lower.5ex\hbox{M}\kern-.125emS}}

\title{Gluon propagator and confinement scenario in Coulomb gauge}

\author{Attilio Cucchieri\address[MCSD]{IFSC S\~ao Paulo University,
          C.P. 369 CEP 13560-970, S\~ao Carlos (SP),
          Brazil}\thanks{Research supported by FAPESP, Brazil
                  (Pro\-ject No.\ 00/05047-5).
                  e-mail address: attilio@if.sc.usp.br}
          and
          Daniel Zwanziger\address{Physics Department, New York
          University, New York, NY 10003, USA}\thanks{Presented talk.
                       Research supported in part by National
                       Science Foundation, grant no.\ PHY-0099393.
                       e-mail address: daniel.zwanziger@nyu.edu}}

\begin{document}

\begin{abstract}
We present numerical results in $SU(2)$ lattice gauge theory for
the instantaneous part of the gluon propagator in Coulomb gauge
$D_{44,{\rm inst}} = V_{\rm coul}(R) \delta(t)$. Data are taken
on lattice volumes $24^4$ and $28^4$ for 7 values of $\beta$ in
the interval
$2.2 \leq \beta \leq 2.8$. The data are confronted with the
confinement scenario in
Coulomb gauge. They are consistent with a linearly rising color-Coulomb
potential $V_{\rm coul}(R)$.
%\vspace{1pc}
\end{abstract}

\maketitle

\section{CONFINEMENT SCENARIO IN COULOMB GAUGE}

A particularly simple confinement scenario \cite{coul} is
available in the minimal Coulomb gauge. It attributes confinement
of color to the long range of the color-Coulomb
potential $V_{\rm coul}(R)$. This quantity is the instantaneous part
of the 4-4 component of the gluon propagator,
$D_{\mu \nu}(x) \equiv  \langle g A_\mu(x) \, g A_\nu(0) \rangle$,
namely $D_{44}({\bf x}, t) = V_{\rm coul}(|{\bf x}|) \delta(t)
+ P({\bf x}, t)$. The
vacuum polarization term $P({\bf x}, t)$ is less singular
than $\delta(t)$ at $t = 0$. Since $A_4$ couples
universally to color charge, the long range of $V_{\rm coul}(R)$
suffices to
confine all color charge. It was conjectured that it is linearly
rising at large
$R$, $V_{\rm coul}(R) \sim - \sigma_{\rm coul}R$.
If an external quark-antiquark pair is
present, the physical potential $V_{\rm W}(R)$ between them may be
extracted from a
Wilson loop. The color-Coulomb potential contributes
the term $ - C V_{\rm coul}(R)$ directly to the Wilson loop,
where $C = (N^2-1)/(2N)$ in SU(N) gauge theory with external
quarks in the
fundamental representation. The minus sign occurs because the
antiquark has
opposite charge to the quark. The vacuum-polarization term is
screening, and one
expects that $V_{\rm W}(R)$ is bounded above by this term
asymptotically at large
$R$, $V_{\rm W}(R) \leq - C V_{\rm coul}(R)$.
If $V_{\rm W}(R)$ is also
linearly rising, $V_{\rm W}(R) \sim \sigma R$, where $\sigma$ is the
conventional
string tension, we get $\sigma \leq C \sigma_{\rm coul}$. If dynamical
quarks are present, the string ``breaks'' at some radius $R_b$, and
the conventional
asymptotic string tension vanishes, $\sigma = 0$. String-breaking is easily
explained in the Coulomb-gauge confinement scenario if $V_{\rm coul}(R)$ is
linearly rising even in the presence of dynamical quarks, as was also
conjectured \cite{coul}. For, if so, it is energetically preferable
to polarize a pair of sea quarks from the vacuum.

Here we report the confrontation of this
scenario with the numerical data for $V_{\rm coul}(R)$ for the case of pure
gluodynamics.

\begin{table*}[t]
\caption{Values of fitting parameters from formula (\ref{eq:fit1}).}
\label{table:1}
\newcommand{\m}{\hphantom{$-$}}
\newcommand{\cc}[1]{\multicolumn{1}{c}{#1}}
\renewcommand{\tabcolsep}{1.35pc} % enlarge column spacing
\renewcommand{\arraystretch}{1.2} % enlarge line spacing
\begin{tabular}{@{}lccccc}
\hline
$\beta$ & $A$ & $B$ & $\Lambda$ & $W^2$ & $\chi^2$/dof \\
\hline
2.2     &  7.38 $\pm$  0.35 &  8.15 $\pm$ 15.58 &
             8.83 $\pm$ 19.34 &  9.91 $\pm$ 19.01 & 0.81 \\
2.3     &  5.81 $\pm$  0.30 & 24.57 $\pm$ 83.41 &
            32.57 $\pm$ 52.59 & 17.26 $\pm$ 27.18 & 0.72 \\
2.4     &  6.60 $\pm$  0.64 &  6.72 $\pm$ 10.15 &
             7.70 $\pm$ 15.90 &  3.99 $\pm$  6.03 & 0.20 \\
2.5     &  7.45 $\pm$  0.47 &  6.00 $\pm$  2.65 &
             6.16 $\pm$  2.58 &  3.06 $\pm$  1.26 & 0.05 \\
2.6     &  8.23 $\pm$  1.48 &  8.61 $\pm$  1.68 &
             4.53 $\pm$  1.93 &  3.84 $\pm$  0.34 & 0.07 \\
2.7     & 11.30 $\pm$  0.81 &  7.73 $\pm$  0.47 &
             6.87 $\pm$  0.67 &  3.64 $\pm$  0.15 & 0.07 \\
2.8     &  8.18 $\pm$  2.85 & 10.38 $\pm$  0.46 &
             0.58 $\pm$  0.08 &  0.06 $\pm$  0.05 & 0.19 \\
\hline
\end{tabular}\\[2pt]
\end{table*}

\subsection{Relation of $V_{\rm coul}$ to $\alpha_s$}

We may identify $V_{\rm coul}(R)$ with the phenomenological potential
that is the starting point for QCD bound state calculations
\cite{szcz}.
The identification of a phenomenological potential with the
instantaneous part of
the gluon propagator, a fundamental
quantity in the gauge theory, is possible because, remarkably,
$V_{\rm coul}(R)$ is a renormalization-group
invariant, and thus scheme-independent, so it is independent of the
cut-off $\Lambda$ and of the renormalization mass $\mu$. This follows from
the non-renormalization of $gA_4$, as expressed by the identity
$g_{(0)}A_4^{(0)} = g_{(r)}A_4^{(r)}$, where $0$ and $r$ refer to
unrenormalized and renormalized quantities in the Coulomb gauge
\cite{coul}. This identity has no direct analog in a Lorentz-covariant
gauge. Because of the scheme-independence of $V_{\rm coul}(R)$, its
Fourier transform $\widetilde{V}_{\rm coul}({\bf k})$ provides a
scheme-independent definition for the running
coupling constant of QCD,
${\bf k}^2\widetilde{V}({\bf k})
= x_0 \ g_{\rm coul}^2(|{\bf k}|)$, and of
$\alpha_s \equiv { {g^2({\bf k}/\Lambda_{\rm coul})} \over {4\pi}}$. Here
$x_0 = { {12N} \over {11N - 2N_f} }$, and $\Lambda_{\rm coul}$ is a finite
QCD mass scale \cite{rgcoul}.

\section{NUMERICAL STUDY OF $V_{\rm coul}$}

\subsection{Method}

We have previously studied \cite{cuzwns}
both space- and time-components of the gluon
propagator, $D_{ij}^{\rm tr}({\bf k})$ and
$D_{44}({\bf k}) = \tilde{V}_{\rm coul}({\bf k})$
at equal time, at $\beta = 2.2$, on various lattice volumes,
$14^4$ to $30^4$, in the minimal Coulomb gauge.
This gauge is achieved numerically by (i) maximizing
$\sum_{x,i=1}^3 {\rm Tr} {^gU}_{x,i}$ with respect to all local gauge
transformations $g(x)$,
where the sum is on all horizontal or spatial links,
and then (ii) maximizing $\sum_{x} {\rm Tr}
{^gU}_{x,4}$ with respect to all ${\bf x}$-independent but
$x_4$-dependent gauge transformations $g(x_4)$
where the sum is on all vertical or time-like links. This makes the 3-vector
potential $A_i$, for $i = 1,2,3$ transverse,
$\partial_i A_i = 0$, so $A_i = A_i^{\rm tr}$.

In this gauge, the horizontal link variables $U_{x,i}$, for $i = 1,2,3$
are as close to the identity as possible, but the vertical variables $U_{x,4}$
are much further from the identity. Not surprisingly, we found that, whereas
$D_{ij}^{\rm tr}({\bf k})$ gave values that could be reasonably
extrapolated to
the continuum, this was not true for
$D_{44}({\bf k})$. In order to remedy this, in the present study we
extended our
investigation of $D_{44}({\bf k})$ to $\beta = 2.2, 2.3, \ldots,
2.8$,
on lattice volumes $24^4$ and $28^4$. We have also determined
$V_{\rm coul}(R)$ numerically by 2 quite different methods.
Method I relies on the standard formula
$D_{44}({\bf x}, t) = V_{\rm coul}(|{\bf x}|)\delta(t) +  P({\bf x}, t)$.
Method II relies on the lattice analog of the continuum formula
that is obtained \cite{rgcoul} by integrating out $A_4$ namely,
$ V(|{\bf x - y}|) =
\langle (M^{-1} (- \nabla^2)M^{-1})|_{\bf x,y} \rangle$.
Here $M({\bf A}) = - {\bf \nabla} \cdot {\bf D}({\bf A})$ is the
3-dimensional Faddeev-Popov operator, and
${\bf D}({\bf A}) = {\bf \nabla} + {\bf A} \times$ is the gauge-covariant
derivative. Method II requires numerically inverting the lattice
Faddeev-Popov matrix to obtain $M^{-1}({\bf A})_{\bf x,y}$, but it has the
advantage that $D_{44}({\bf k})$ is expressed entirely in terms of the
spatial link variables $U_{x,i}$ for $i = 1,2,3$,
that are close to the identity. Moreover, method II involves only
the horizontal link variables that lie within a single time slice, so it is
independent of the gauge fixing (ii) on vertical links. Thus it measures
a truly instantaneous quantity. We found that method I did not exhibit
scaling in the above range of $\beta$, and we
report here only the result of method II.

\begin{figure}[t]
\vspace*{-3.6cm}
\begin{center}
\epsfxsize=0.39\textwidth
\leavevmode\epsffile{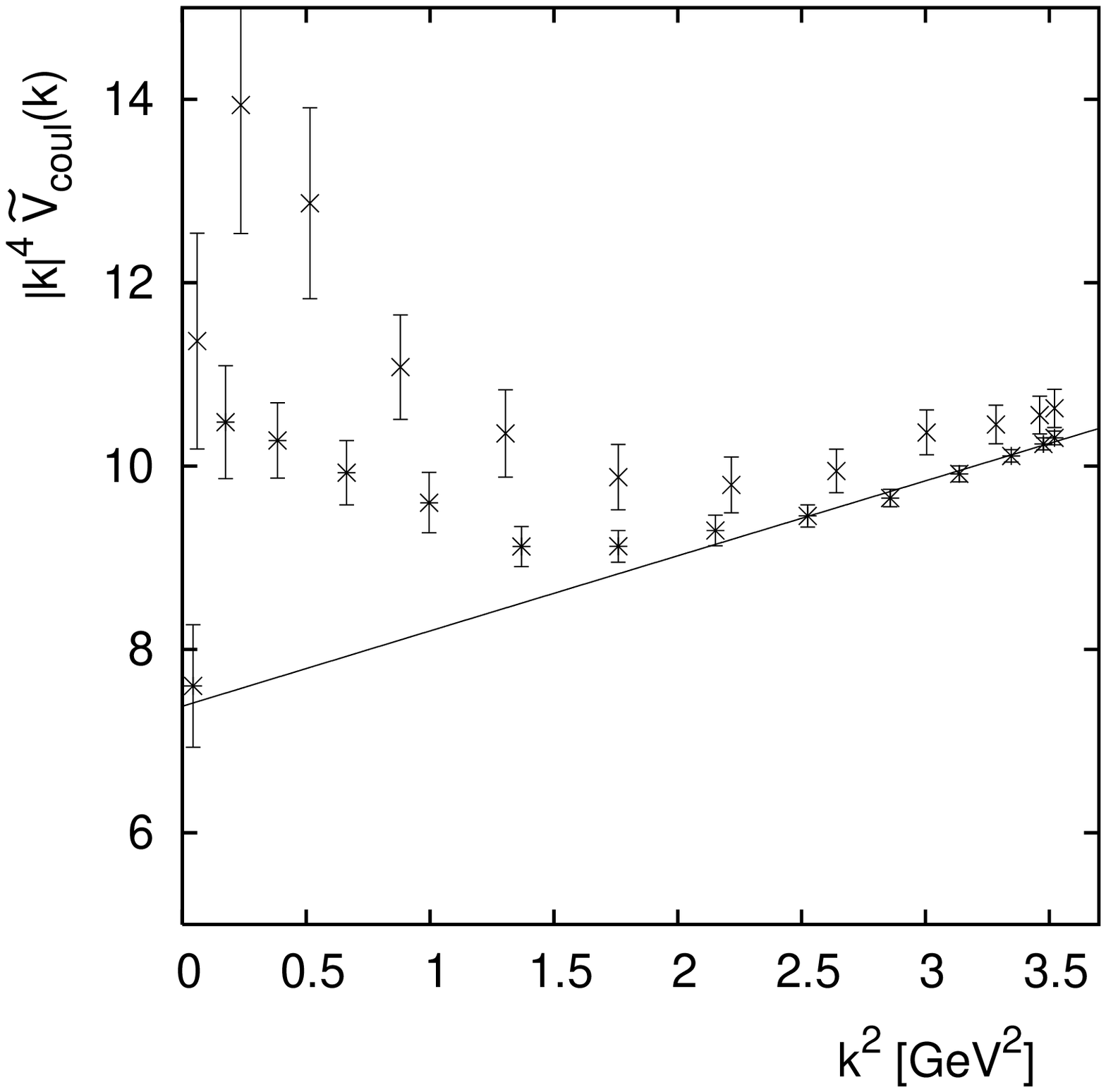}
\end{center}
\begin{center}
\vspace*{-3.3cm}
\epsfxsize=0.39\textwidth
\leavevmode\epsffile{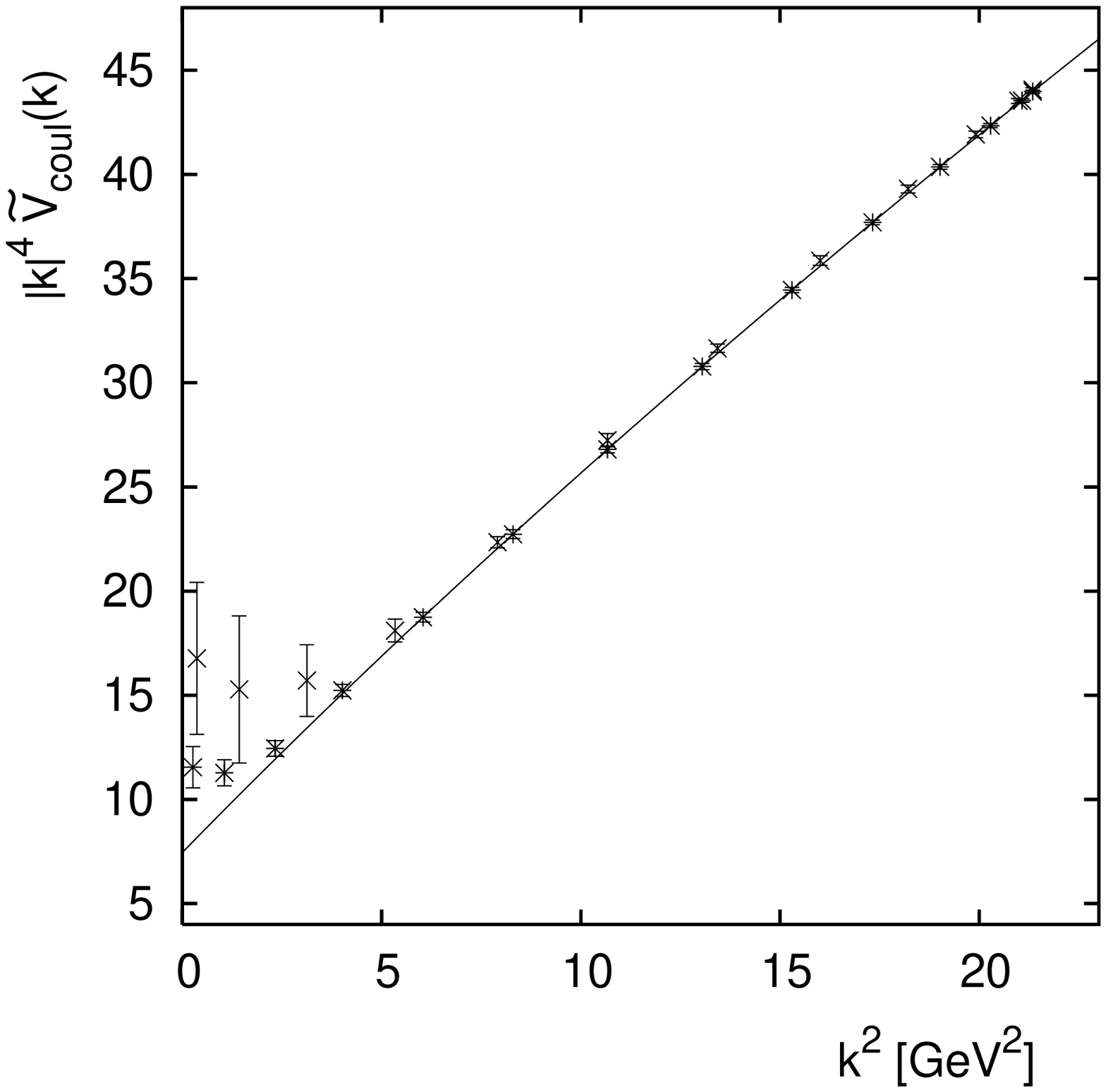}
\end{center}
\begin{center}
\vspace*{-3.3cm}
\epsfxsize=0.39\textwidth
\leavevmode\epsffile{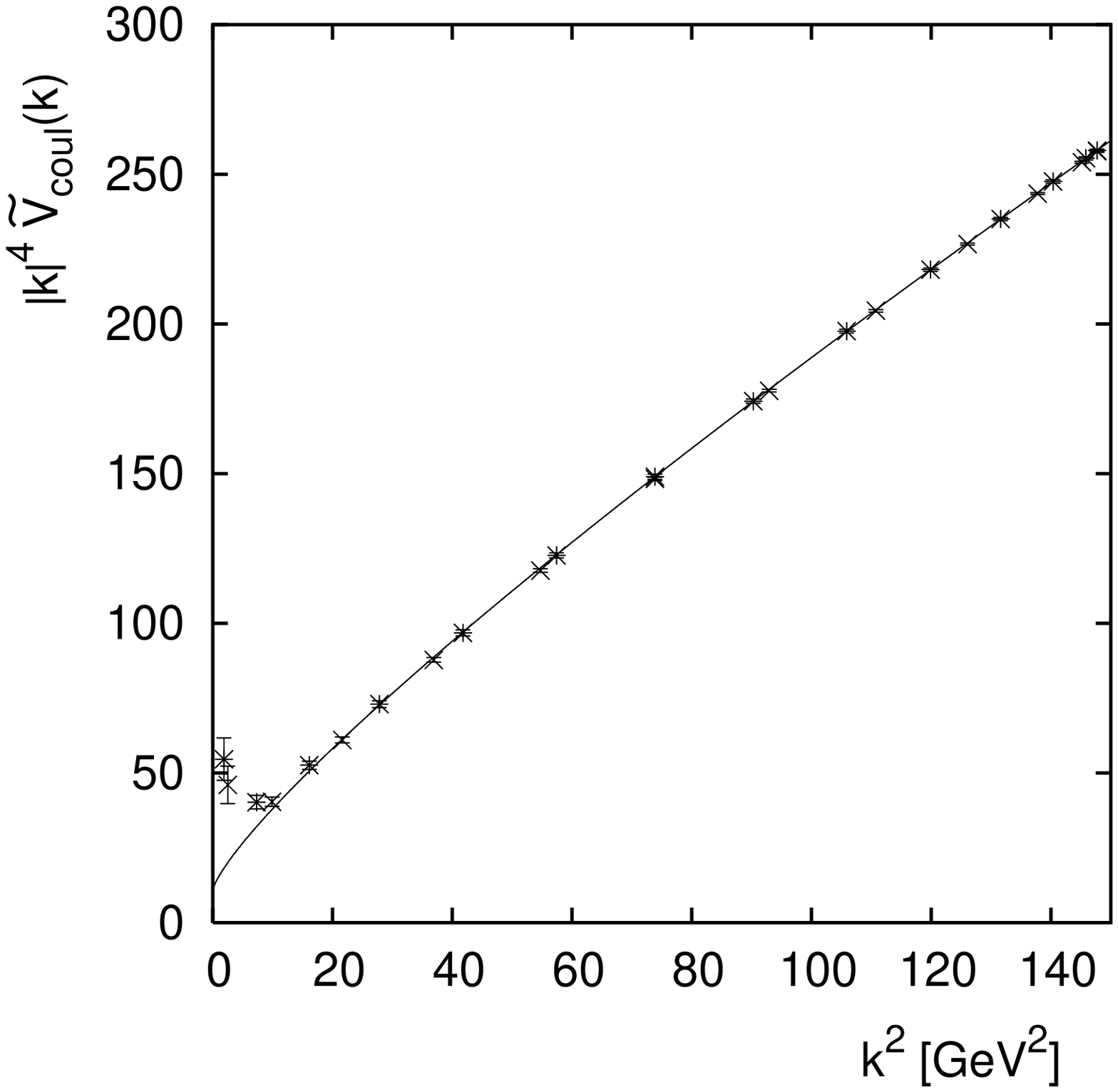}
\vspace*{-0.7cm}
\caption{Fit of $|{\bf k}|^4 \tilde{V}_{\rm coul}({\bf k})$
using eq.\ (\ref{eq:fit1}). Data are
for lattice volumes $V = 24^4$ ($\times$) and
$V = 28^4$ ($\ast$) and $\beta = 2.2, 2.5$ and $2.8$
from top to bottom.}
\label{fig:k4V}
\end{center}
\end{figure}

\subsection{Results}

For each $\beta$ we have 100 configurations on volume
$V = 24^4$ and 50 configurations on volume $V = 28^4$.
Runs have been done on the PC clusters at the
Physics Department of New York University and
at the IFSC of S\~ao Paulo University.

In figure 1 we plotted the results for $\beta =$ 2.2, 2.5, and
2.8 respectively. The horizontal axis measures ${\bf k}^2 = 4 a^{-2}
\sin^2(n\pi/L)$, for
$L = 24, 28$, rescaled to physical units by
setting the physical string tension equal to
$\sigma = (0.44 {\rm GeV})^2$ and using
Table 3 of \cite{avalue},
so the lattice spacing at
$\beta = 2.2, \ldots, 2.8$ is (in ${\rm GeV}^{-1}$)
$a=$ 1.066, 0.839, 0.605, 0.433, 0.309, 0.231, 0.165,
respectively. The vertical axis measures
$|{\bf k}|^4 \tilde{V}_{\rm coul}({\bf k})$ in physical units, so
string tension
may be read off from the vertical intercept (see below).
Finite-volume artifacts
are clearly visible at low momentum. To control these, in our fits we have
dropped those low-momentum points for which appreciably different values are
obtained at volumes $24^4$ and $28^4$ and, for the fit, we used only
data points obtained at $V = 28^4$.

A simple parametrization of $V_{\rm coul}(R)$ would be
$ - V_{\rm coul}(R) = \sigma_{\rm coul} R - c/R$, which has the
Fourier transform
$ \tilde{V}_{\rm coul}({\bf k})
= 8 \pi \sigma_{\rm coul}/|{\bf k}|^4 + 4 \pi c/|{\bf k}|^2$.   We
have used the
fitting formulas
\begin{eqnarray}
|{\bf k}|^4 \tilde{V}_{\rm coul}({\bf k})  \!\!\! & = & \!\!\! A
+ { {B{\bf k}^2} \over {W^2 + \ln(1 +{\bf k}^2/\Lambda^2) } }
\label{eq:fit1} \\
|{\bf k}|^4 \tilde{V}_{\rm coul}({\bf k}) \!\!\! & = & \!\!\!
                                              A + B{\bf k}^2 \\
|{\bf k}|^4 \tilde{V}_{\rm coul}({\bf k}) \!\!\! & = & \!\!\!
                                             A|{\bf k}|^s + B{\bf k}^2
            \; .
\end{eqnarray}
Fit (\ref{eq:fit1})
has the asymptotic behavior at large ${\bf k}$, consistent with
the 1-loop $\beta$-function.
We report in Table~\ref{table:1} the values of the parameters
from fit~(\ref{eq:fit1}).

The most striking feature of the data is the finite intercepts
$A$. This
is consistent with a finite string tension
$\sigma_{\rm coul} = A/(8 \pi)$.  It scales rather nicely,
varying from
$A = 7.38  \pm  0.35 \ {\rm GeV}^2$ at $\beta = 2.2$ to
$A = 8.18  \pm  2.85 \ {\rm GeV}^2$ at $\beta = 2.8$,
but with considerable variation in between.  From the lowest and
highest values,
$A = 5.8$ and
$A = 11.3$, we get respectively
$\sigma_{\rm coul} = (0.48 {\rm GeV})^2$
and $\sigma_{\rm coul} = (0.67 {\rm GeV})^2$.  The inequality
$\sigma \leq (3/4)\sigma_{\rm coul}$ for SU(2), with
$\sigma = (0.44{\rm GeV})^2$, reads for these values:
$0.44 \leq 0.42(2)$ and $0.44 \leq 0.58(3)$.  We appear to be at or near
saturation.


\begin{thebibliography}{9}

\bibitem{coul} D. Zwanziger,
                %{\it Renormalization in the Coulomb gauge and
                %order parameter for confinement in QCD},
                Nucl.\ Phys.\ {\bf B518} (1998) 237.

\bibitem{szcz} A. Szczepaniak et al.,
                % E. S. Swanson, C.-R. Ji and S. R. Cotanch,
                Phys.\ Rev.\ Lett.\ {\bf 76} (1996) 2011;
                D. G. Robertson et al.,
                % E. S. Swanson, A. P. Szczepaniak, C.-R.
                % Ji and S. R. Cotanch,
                %{\it Renormalized effective QCD Hamiltonian: gluonic sector}
                Phys.\ Rev.\ {\bf D59} (1999) 074019;
                A. Szczepaniak and E. S. Swanson,
                %{\it Coulomb gauge QCD, confinement and the constituent
                % representation},
                hep-ph/0107078.

\bibitem{rgcoul} A. Cucchieri and D. Zwanziger,
                  Phys. Rev. {\bf D65} (2001) 014002.

\bibitem{cuzwns} A. Cucchieri and D. Zwanziger, Phys. Rev. {\bf D65}
                  (2001) 014001;
                  Phys. Letts. {\bf B524} (2002) 123.

\bibitem{avalue} J. Fingberg et al.,
                  % Urs M. Heller, F. Karsch
                  Nucl.\ Phys.\ {\bf B392} (1993) 493.

\end{thebibliography}
\end{document}